\documentclass[journal]{IEEEtran}
\IEEEoverridecommandlockouts  
\usepackage[compress]{cite}
\usepackage[cmex10]{amsmath}
\usepackage{amssymb}
\usepackage{array}
\usepackage{multicol}
\usepackage{amsfonts}
\usepackage{graphicx}
\usepackage{subfigure}
\usepackage{verbatim}
\usepackage[named]{algonew}
\usepackage{algorithmic}
\usepackage{setspace}
\makeatletter

\newcommand{\Rmnum}[1]{\expandafter\@slowromancap\romannumeral #1@}
\makeatother

\title{Transmission Mode Selection for Downlink Coordinated Multi-Point Systems}
\author{{ Qian Zhang}, {\em Student Member, IEEE} and {Chenyang Yang}, {\em Senior Member, IEEE}
\thanks{Copyright (c) 2012 IEEE. Personal use of this material is permitted.
However, permission to use this material for any other purposes must
be obtained from the IEEE by sending a request to
pubs-permissions@ieee.org.

This work was supported by the National Key project of Next
Generation Wideband Wireless Communication Network (2011ZX03003-001)
and the International S\&T Cooperation Program of China (ISCP)
(2008DFA12100).

Q. Zhang and C. Yang are with the School of Electronics and
Information Engineering, Beihang University, Beijing 100191, China.
(E-mail: qianzhang@ee.buaa.edu.cn, cyyang@buaa.edu.cn)} }

\begin{document}

\maketitle

\begin{abstract} \vspace{-0.1cm}
Coordinated multi-point (CoMP) transmission has been widely
recognized as a spectrally efficient technique in future cellular
systems. To exploit the abundant spatial resources provided by the
cooperating base stations, however, considerable training overhead
is required to acquire the channel information. To avoid the extra
overhead outweighing the cooperative gain, we propose a method that
allows each user to select transmission mode between coherent CoMP
and Non-CoMP. We first analyze the average throughput of each user
under CoMP and Non-CoMP transmission after taking into account the
downlink training overhead. A closed-form mode selection rule is
then developed, which depends on the user location and system
settings, i.e, the number of cooperating base stations and transmit
antennas, training overhead and cell-edge signal to noise ratio.
Simulation results show that the proposed downlink transmission mode
selection method achieves higher throughput than CoMP for
cell-center users and than Non-CoMP for cell-edge users after
accounting for the overhead. As a by-product, the backhaul load is
also reduced significantly.
\end{abstract}
\vspace{-0.0cm}
\begin{keywords}\vspace{-0.1cm}
Coordinated multi-point transmission, transmission mode selection,
training overhead
\end{keywords}
\vspace{-0.0cm}
\section{Introduction}
\label{sec:intro}

Recently, base station (BS) cooperative transmission, known as
coordinated multi-point (CoMP) in 3GPP long-term evolution
(LTE)-advanced, has been widely recognized as a promising technique
to enhance throughput by avoiding inter-cell interference (ICI)
especially for cell-edge users
\cite{Karakayali2006,Tolli08,Molisch2008}.

The BS cooperative strategies  can be roughly divided into CoMP
joint processing (CoMP-JP) and coordinated beamforming (CoMP-CB),
depending on the information exchanged among the BSs. CoMP-JP can
exploit the abundant spatial resources provided by the cooperating
BSs with joint multi-user multi-input-multi-output (MU-MIMO)
precoding, where both data and channel state information (CSI) need
to be shared \cite{Karakayali2006}. By contrast, CoMP-CB avoids ICI
by using individual precoding at each BS, where only CSI is shared
\cite{Huang2009}. Since sharing CSI requires much lower capacity
than sharing data \cite{Huang2009}, CoMP-CB needs much lower
backhaul capacity than CoMP-JP. In \cite{Jun2010,Nima2010,CR2010},
the performance of CoMP-JP, CoMP-CB and Non-CoMP systems was
compared. The results show that when the cell-edge signal to noise
ratio (SNR) is high, CoMP-CB is superior to Non-CoMP \cite{Jun2010},
and even outperforms CoMP-JP if the backhaul capacity is low
\cite{Nima2010}. When the number of cooperative BSs or the number of
antennas at each BS is large, CoMP-JP has no throughput gain over
Non-CoMP after accounting for the training overhead to assist
channel estimation \cite{CR2010}.

A key factor differentiating the two categories of CoMP is the
backhaul. If the backhaul has infinite-capacity and zero latency,
CoMP-JP is more spectrally efficient than CoMP-CB. This is true even
when the training overhead is taken into account, because the two
CoMP strategies need comparable overhead. Note that although the
backhaul links in existing cellular systems have much lower capacity
than CoMP-JP requires \cite{Huang2009}, there are no technical
challenges to upgrade the backhaul with high speed optical fiber.
When the backhaul links are perfect, however, CoMP-JP may not always
be superior to Non-CoMP as expected \cite{CR2010}. This is because
the performance of CoMP transmission largely depends on the users'
location, i.e., cell-edge users will benefit more from cooperative
transmission than cell-center users, and the throughput gain for the
cell-center users may be counteracted by the extra training overhead
in practice.

In this paper, we strive to mitigate the adverse effect of training
overhead on the downlink throughput of CoMP system by switching
between CoMP-JP and Non-CoMP transmission modes. Transmission mode
selection has been extensively studied in single cell MIMO systems,
e.g., \cite{Runhua08,Schellmann2010}. By switching the mode based on
channel conditions between transmitting single and multiple data
streams \cite{Runhua08} or between applying statistical beamforming
and spatial multiplexing \cite{Schellmann2010}, either spectral
efficiency can be increased or transmission reliability can be
improved. Yet there are few works in the literature addressing the
transmission mode selection in multi-cell systems. In
\cite{Jun2010}, the authors suggested to switch the transmission
strategy between CoMP-CB and Non-CoMP at the BSs side to maximize
the sum rate, depending on whether the user is noise- or
interference-limited.

In contrast to the transmission mode selection in single cell
systems that primarily depends on the channel conditions, mode
selection in CoMP depends on the backhaul and overhead as well. In
this paper, we consider the backhaul with unlimited-capacity. To
achieve a trade-off between the cooperative gain and the training
overhead, we develop a method for each user to select either CoMP-JP
or Non-CoMP transmission mode. To avoid introducing additional
control overhead, we select transmission mode based on the
statistical channel information at the user side.

\section{System model}
\label{sec:format}

Consider a cooperative cluster consisting of $B$ BSs each equipped
with $N_t$ antennas. $K$ users each with single antenna are located
in each cell, and each user treats the closest BS as its local BS.
The assumption of single antenna users is for simplicity and does
not preclude applying the proposed method to multiple antenna users.
Denote $\mathbf{g}_{iu}\in\mathbb{C}^{N_t\times 1}$ as the
small-scale fading channel vector between BS $i$ and user $u$, each
entry of which is complex Gaussian random variable with unit
variance, and all the channel vectors are assumed as independent and
identically distributed (\emph{i.i.d.}). $\mathbf{h}_u=
[\alpha_{1u}\mathbf{g}_{1u}^H,\cdots,\alpha_{Bu}\mathbf{g}_{Bu}^H]^H\in\mathbb{C}^{BN_t\times1}$
represents the global channel vector of user $u$, where
$\alpha_{iu}$ is the large-scale fading channel gain from BS $i$ to
the user, which includes path loss and shadowing. For simplicity, we
refer CoMP-JP as CoMP in the rest of the paper.

Under CoMP transmission mode, we consider that the $B$ BSs are
connected with a central unit (CU) via backhaul links of unlimited
capacity and zero latency. After collecting CSI from each BS, the CU
selects multiple users from the user pool in the $B$ cells and then
computes the global precoding vectors for the co-scheduled users.
Then it sends the precoded data to the $B$ BSs, who jointly transmit
to the active users. Consider that the number of total active users
jointly served by the $B$ BSs is $BM$, where $M\leq K$ and $M\leq
N_t$. $BM-1$ partner users are co-scheduled with user $u$ to share
the same time-frequency resource with it.

The received signal of user $u$ under CoMP transmission mode is
given by \vspace{-0.2cm}
\begin{equation} \label{E:TRmodelCoMP}
  y_u^{\mathrm{C}} = \mathbf{h}_{u}^H \mathbf{v}_{u}x_u +
  \sum_{j=1}^{BM-1}\mathbf{h}_{u}^H \mathbf{v}_{q_j}x_{q_j} +
  z_u, \vspace{-0.3cm}
\end{equation}
where $(\cdot)^H$ is the conjugate transpose of a vector or matrix;
$x_u, x_{q_j}$ are the data intended to user $u$ and user $q_j$ that
have unit average energy, i.e.,
$\mathbf{E}\{||x_u||^2\}=\mathbf{E}\{||x_{q_j}||^2\}=1$; $z_u$
denotes the noise at user $u$, which is a white Gaussian random
variable with zero mean and variance $\sigma^2$;
$\mathbf{v}_{u},\mathbf{v}_{q_j}\in\mathbb{C}^{BN_t\times 1}$ are
the global precoding vectors for user $u$ and user $q_j$;
$\mathrm{IUI}_{q_j}^{\mathrm{C}}\triangleq\mathbf{h}_{u}^H
\mathbf{v}_{q_j}x_{q_j}$ represents the inter-user interference
(IUI) from user $q_j$ to user $u$, and $\triangleq$ means
definition.

We consider zero-forcing (ZF) precoder for downlink MU-MIMO
transmission, which is a low-complexity yet asymptotically optimal
precoder \cite{Yoo06-SUS}. The precoding matrix can be expressed as
$\mathbf{V}=\mathbf{H}(\mathbf{H}^H\mathbf{H})^{-1}\mathbf{P}$,
where $\mathbf{V} = [\mathbf{v}_u,
\mathbf{v}_{q_1},\dots,\mathbf{v}_{q_{BM-1}}]$,
$\mathbf{H}=[\mathbf{h}_u,\mathbf{h}_{q_1},\dots,\mathbf{h}_{q_{BM-1}}]$
is the channel matrix of the $BM$ active users, and
$\mathbf{P}=\mathrm{diag}\{\sqrt{p_{u}^{\mathrm{C}}},\sqrt{p_{q_1}^{\mathrm{C}}},
\dots,\sqrt{p_{q_{BM-1}}^{\mathrm{C}}}\}$ represents the power
allocation matrix. Then the received signal to noise plus
interference ratio (SINR) of user $u$ is \vspace{-0.15cm}
\begin{equation} \label{E:GammaC}
  \gamma_u^{\mathrm{C}} = \frac{p_{u}^{\mathrm{C}}}
  {\sigma^2}. \vspace{-0.05cm}
\end{equation}

Under Non-CoMP transmission mode, each BS selects multiple users
from the $K$ users located in its serving cell. Subsequently, each
BS serves these active users with ZF precoding, and each user
receives the desired signal from its local BS suffering ICI from
other BSs. In order to serve the same number of users in the whole
cluster as in the CoMP transmission mode, we consider that each BS
serves $M$ active users in the Non-CoMP transmission mode.

The received signal of user $u$ in cell $b$ under this mode is given
by \vspace{-0.3cm}
\begin{align} \nonumber
  y_u^{\mathrm{NC}} &=
  \sum_{i=1}^B\alpha_{iu}\mathbf{g}_{iu}^H\mathbf{W}_i\mathbf{x}_i +
  z_u \ =\ \alpha_{bu}\mathbf{g}_{bu}^H \mathbf{w}_{bu}x_u
   + \\ \label{E:TRmodel}
  &
  \sum_{j=1}^{M-1}\alpha_{bu}\mathbf{g}_{bu}^H
  \mathbf{w}_{bs_j}x_{s_j}
  +
  \sum_{i=1,i\neq
  b}^B\alpha_{iu}\mathbf{g}_{iu}^H\mathbf{W}_i\mathbf{x}_i
  +z_u, \vspace{-0.3cm}
\end{align}
where $\mathbf{x}_i\in\mathbb{C}^{M\times 1}$ is the data vector at
BS $i$ for its $M$ active users; $\mathbf{W}_i =
[\mathbf{w}_{iu},\mathbf{w}_{is_1}, \cdots,\mathbf{w}_{is_{M-1}}]
\in\mathbb{C}^{N_t\times M}$ is the precoding matrix at BS $i$,
whose columns represent the precoding vectors for the $M$ active
users; $\mathbf{g}_{bu}$ is the channel vector of user $u$ in cell
$b$,
$\mathrm{IUI}_{bs_j}^{\mathrm{NC}}\triangleq\alpha_{bu}\mathbf{g}_{bu}^H\mathbf{w}_{bs_j}x_{s_j}$
represents the IUI from user $s_j$ to user $u$, and
$\mathrm{ICI}_{iu}\triangleq\alpha_{iu}
\mathbf{g}_{iu}^H\mathbf{W}_i\mathbf{x}_i$, $i\neq b$, represents
the ICI from BS $i$ to user $u$.

The ZF precoding matrix at BS $b$ is $\mathbf{W}_b
=\mathbf{G}_b(\mathbf{G}_b^H\mathbf{G}_b)^{-1}\mathbf{P}_b$, where
$\mathbf{G}_b=[\mathbf{g}_{bu},\mathbf{g}_{bs_1},\dots,\mathbf{g}_{bs_{M-1}}]$
is the channel matrix of the $M$ active users in cell $b$, and
$\mathbf{P}_b=\mathrm{diag}\{\sqrt{p_{u}^{\mathrm{NC}}},\
\sqrt{p_{s_{1}}^{\mathrm{NC}}},\dots,\sqrt{p_{s_{M-1}}^{\mathrm{NC}}}\}$
is the power allocation matrix. By assuming that all the ICIs are
uncorrelated Gaussian noises, and further considering the average
interference derived in \cite{ZY112},
$\mathbf{E}\{|\mathrm{ICI}_{iu}|^2\}=P\alpha_{iu}^2$, the received
SINR of user $u$ under Non-CoMP transmission mode is obtained as
\vspace{-0.0cm}
\begin{equation} \label{E:GammaNC}
  \gamma_u^{\mathrm{NC}} = \frac{\alpha_{bu}^2p_{u}^{\mathrm{NC}}}
  {\sum_{i\neq b}^{B}\mathbf{E}|\mathrm{ICI}_{iu}|^2
  +\sigma^2}= \frac{\alpha_{bu}^2p_{u}^{\mathrm{NC}}}
  {P\sum_{i\neq b}^{B}\alpha_{iu}^2
  +\sigma^2}, \vspace{-0.01cm}
\end{equation}
which can serve as a lower bound of the SINR.

\section{Transmission Mode Selection Considering Training Overhead}

In this section, we design a method for downlink transmission mode
selection accounting for the impact of downlink training overhead.
By choosing a transmission mode between CoMP and Non-CoMP, each user
can achieve its maximal net throughput. As a result, the system can
attain a higher overall throughput. \vspace{-0.4cm}

\subsection{Net User Throughput Considering Training Overhead}
To facilitate downlink MU-MIMO precoding, the CSI of scheduled users
should be available at the CU. In frequency division duplexing (FDD)
systems, the CSI is first estimated at the user side via downlink
training then is obtained via uplink feedback with various
techniques such as limited feedback. In time division duplexing
(TDD) systems, the CSI is estimated via uplink training by
exploiting channel reciprocity, or obtained with limited feedback
when the reciprocity does not hold due to antenna calibration errors
\cite{Caire11}. Since it is not proper to simply count the uplink
overhead into the downlink throughput \cite{Caire11}, we only
consider downlink training overhead as in \cite{CR2010}. Although
channel estimation errors or quantization errors have large impact
on the throughput of MU-MIMO, the impacts on CoMP and Non-CoMP are
similar. Limited feedback strategies for CoMP is an ongoing research
topic \cite{HXY2011}, which will not be explored here. To highlight
the impact of the overhead, we suppose that perfect CSI can be
obtained, and leave the imperfect CSI issues for future work.

Except for the downlink training for CSI feedback (i.e., the common
pilots in the context of LTE-Advanced), dedicated pilots are also
required during downlink transmission to assist each user to
estimate an equivalent channel after precoding for data detection.
The overhead of the common pilots and that of the dedicated pilots
are respectively in proportion to the number of BS antennas and that
of data streams, and both occupy downlink time or frequency
resources.

Specifically, consider a block fading channel with $C=T_cW_c$
channel uses in a coherence block, where $T_c$ and $W_c$ are
respectively the channel coherence time and coherence bandwidth
\cite{Caire11}. In Non-CoMP systems, suppose that $C_c$ channel uses
are employed for common pilots of each antenna and $C_d$ channel
uses are employed for dedicated pilots of each data stream. Then the
downlink training overhead is
$v^{\mathbf{NC}}=\frac{N_tC_c+N_rC_d}{C}$, where $N_r$ is the number
of data streams and is equal to one in this paper since we assume
single antenna users. In CoMP systems, the overhead can be expressed
as $v^{\mathbf{C}}=\frac{\beta BN_tC_c+\epsilon N_rC_d}{C}$, where
$\beta \leq 1$ indicates that CoMP systems should use sparser common
pilots\footnote{Sparser pilots will lead to less accurate channel
estimation, which is not modeled here.} than Non-CoMP systems
otherwise the overhead will be too large, and $\epsilon \geq 1$
indicates that more dedicated pilots are required to enhance the
orthogonality among different users, as suggested in \cite{3GPP-RS}
and the references therein. The scaling factor of $B$ in
$v^{\mathbf{C}}$ is because the inter-cell common pilots are
orthogonal, as suggested in \cite{3GPP-RS2}.\footnote{An example of
the structure of inter-cell common pilots can be found in a 3GPP
proposal \cite{3GPP-CRS}, which is mentioned in \cite{3GPP-RS2}.}

After taking into account the downlink training overhead, the net
downlink throughputs of user $u$ under CoMP and Non-CoMP
transmission can be respectively expressed as \cite{CR2010}
\vspace{-0.05cm}
\begin{align} \label{E:RateO}
  &R_u^{\mathrm{C}} = (1 - v^{\mathbf{C}}) \log_2 \left( 1+
  \gamma_u^{\mathrm{C}}\right),\\
    &R_u^{\mathrm{NC}} = (1 - v^{\mathbf{NC}}) \log_2 \left( 1+ \gamma_u^{\mathrm{NC}}\right)
\label{E:RateO2}.\ \ \
\end{align}
It shows that the net throughputs decrease with the overhead
linearly but increase with SINR in log scale. Therefore the
throughput gain of CoMP may be counteracted by its training
overhead, although $\gamma_u^{\mathrm{C}}> \gamma_u^{\mathrm{NC}}$.
This motivates the transmission mode selection from the view of each
user. \vspace{-0.5cm}

\subsection{Transmission Mode Selection}
To maximize the overall net throughput of the system, we should
allow each user, say user $u$, to select CoMP transmission when
$R_u^{\mathrm{C}}>R_u^{\mathrm{NC}}$ but Non-CoMP when
$R_u^{\mathrm{C}}<R_u^{\mathrm{NC}}$. However, such a transmission
mode selection is dynamic because $R_u^{\mathrm{C}}$ and
$R_u^{\mathrm{NC}}$ depend on small-scale fading channels, which can
achieve better performance but will induce large signalling overhead
and high protocol complexity. In practice, semi-dynamic mode
selection based on average channel gains is more preferable.
Therefore, we consider a rule for selecting CoMP transmission mode
as follows, \vspace{-0.1cm}
\begin{equation} \label{E:MSrule}
  \mathbf{E}\{{R}_u^{\mathrm{C}}\}>\mathbf{E}\{{R}_u^{\mathrm{NC}}\}, \vspace{-0.1cm}
\end{equation}
where $\mathbf{E}\{\cdot\}$ is the expectation over small-scale
fading channels.

From (\ref{E:RateO}) and (\ref{E:RateO2}), we obtain upper bounds of
$\mathbf{E}\{{R}_u^{\mathrm{C}}\}$ and
$\mathbf{E}\{{R}_u^{\mathrm{NC}}\}$ by using Jensen's inequality,
\vspace{-0.05cm}
\begin{align}  \label{E:rateAppro}
  &\mathbf{E}\{R^{\mathrm{C}}_{u}\}
  \leq (1 - v^{\mathbf{C}})\log_2\left(1+
  \mathbf{E}\{\gamma_{u}^{\mathrm{C}}\}\right)\triangleq\mathbf{E}\{R^{\mathrm{C}}_{u}\}^{ub},
 \\ \label{E:rateAppro1}
  & \mathbf{E}\{R^{\mathrm{NC}}_{u}\}
  \leq (1 - v^{\mathbf{NC}})\log_2\left(1+
  \mathbf{E}\{\gamma_{u}^{\mathrm{NC}}\}\right)\triangleq\mathbf{E}\{R^{\mathrm{NC}}_{u}\}^{ub}. \vspace{-0.2cm}
\end{align}
In later simulations, we will show that the mode selection using
these two upper bounds instead of the true values of the average
throughput has negligible impact on the system performance.

The expressions of the upper bounds can also be written in the
following forms \vspace{-0.15cm}
\begin{align} \label{E:RateUB1}
  &\mathbf{E}\{R^{\mathrm{C}}_{u}\}^{ub} =
  \log_2(1+{\eta}^{\mathrm{C}}\mathbf{E}\{\gamma_{u}^{\mathrm{C}}\}),
  \\ \label{E:RateUB2}
   & \mathbf{E}\{R^{\mathrm{NC}}_{u}\}^{ub} =
  \log_2(1+{\eta}^{\mathrm{NC}}\mathbf{E}\{\gamma_{u}^{\mathrm{NC}}\}), \vspace{-0.3cm}
\end{align} \vspace{-0.3cm}
where \vspace{-0.1cm}
\begin{align} \nonumber
  &\eta^{\mathrm{C}}\triangleq\frac{\left(\mathbf{E}\{\gamma_{u}^{\mathrm{C}}\}+1\right)^{1 -
  v^{\mathbf{C}}}
  -1}{\mathbf{E}\{\gamma_{u}^{\mathrm{C}}\}} \ \ \ \mathrm{and} \ \ \\
  \label{E:eta1}
  &\eta^{\mathrm{NC}}\triangleq\frac{\left(\mathbf{E}\{\gamma_{u}^{\mathrm{NC}}\}+1\right)^{1 -
  v^{\mathbf{NC}}}
  -1}{\mathbf{E}\{\gamma_{u}^{\mathrm{NC}}\}} \vspace{-0.3cm}
\end{align}
respectively reflect the impact of the overhead on the SINR under
CoMP and Non-CoMP transmission mode, and
$0<\eta^{\mathrm{C}}<\eta^{\mathrm{NC}}<1$. Then the decision rule
in (\ref{E:MSrule}) can be derived as \vspace{-0.1cm}
\begin{equation} \label{E:LScompare}
  {\eta}^{\mathrm{C}}\cdot\mathbf{E}\{\gamma_u^{\mathrm{C}}\} >
  {\eta}^{\mathrm{NC}}\cdot\mathbf{E}\{\gamma_u^{\mathrm{NC}}\}. \vspace{-0.15cm}
\end{equation}

To obtain a closed-form decision rule for selecting transmission
mode, we will derive the expressions of
$\mathbf{E}\{\gamma_u^{\mathrm{C}}\}$ and
$\mathbf{E}\{\gamma_u^{\mathrm{NC}}\}$ in the following.

\subsubsection{Average SINR under CoMP transmission}
Denote the transmit power at each BS as $P$. For analytical
tractability, we assume that under CoMP transmission mode the sum
power of all BSs, i.e., $BP$, is equally allocated to the $BM$
active users. This corresponds to per-user power constraint (PUPC)
as in \cite{Molisch2008}, whose performance approaches that of
per-BS power constraint (PBPC) when each cell has a large number of
users \cite{ZhangJun2009}. Then we have \vspace{-0.15cm}
\begin{equation}\label{E-PC}
  p_{u}^{\mathrm{C}} = \frac{BP}{BM\left[
  (\mathbf{H}^{H}\mathbf{H})^{-1}\right]_{1,1}}
  = \frac{P}{M\left[
  (\mathbf{H}^{H}\mathbf{H})^{-1}\right]_{1,1}}, \vspace{-0.2cm}
\end{equation}
where $\left[\cdot\right]_{k,k}$ denotes the element on the $k$th
row and $k$th column of a matrix.

Define
$\theta_u^{\mathrm{C}}\triangleq\angle(\mathbf{h}_{u},\overline{\mathbf{H}})$
as the angle between the channel of user $u$ and a subspace spanned
by the channels of its $BM-1$ co-scheduled users under CoMP
transmission, where
$\overline{\mathbf{H}}=[\mathbf{h}_{q_1},\dots,\mathbf{h}_{q_{BM-1}}]$.
Then we can derive that \vspace{-0.2cm}
\begin{equation} \label{E:AngleC}
  \frac{1}{[(\mathbf{H}^H\mathbf{H})^{-1}]_{1,1}} =
  \left|\mathbf{h}^H_{u}\left(\mathbf{I}-\overline{\mathbf{H}}(\overline{\mathbf{H}}^H
  \overline{\mathbf{H}})^{-1}\overline{\mathbf{H}}^H\right)\right|^2
  = |\mathbf{h}_{u}|^2\sin^2(\theta_u^{\mathrm{C}}). \vspace{-0.0cm}
\end{equation}

Substituting (\ref{E-PC}) and (\ref{E:AngleC}) into
(\ref{E:GammaC}), the received SINR of user $u$ can be obtained as
\vspace{-0.15cm}
\begin{equation} \label{E:GcEPA}
  \gamma_{u}^{\mathrm{C}}  =
  \frac{P}{M\sigma^2[(\mathbf{H}^{H}\mathbf{H})^{-1}]_{1,1}}
  =\frac{P\ |\mathbf{h}_{u}|^2\delta_{BM-1}} {M\sigma^2}, \vspace{-0.2cm}
\end{equation}
where $\delta_{BM-1}\triangleq\sin^2(\theta_u^{\mathrm{C}})$, whose
value is between 0 and 1, a larger value of it indicates a better
orthogonality between user $u$ and its co-scheduled users, which
leads to a larger value of $\gamma_{u}^{\mathrm{C}}$.

From (\ref{E:GcEPA}), the average SINR of user $u$ is
$\mathbf{E}\{\gamma_{u}^{\mathrm{C}}\} =\frac{P} {M\sigma^2}
\mathbf{E}\{|\mathbf{h}_{u}|^2\delta_{BM-1}\}$. The random variables
$|\mathbf{h}_{u}|^2$ and $\delta_{BM-1}$ are in general mutually
dependent. Nonetheless, if all co-scheduled users have equal
large-scale channel gains from all the $B$ BSs\footnote{Note that
the global channels of cell-edge users do not necessarily exhibit
such a statistic. We refer to the users with such kind of channels
as ``cell-edge users" because they behave like cell-edge users.},
their global channels become
 $\mathbf{h}_j=\alpha
[\mathbf{g}_{1j}^H,\cdots,\mathbf{g}_{Bj}^H]^H, j=1,\cdots,BM$ and
are \textit{i.i.d.}, because $\mathbf{g}_{ij}$ was assumed
\textit{i.i.d.}. This corresponds to the worst case in CoMP systems,
since each ``cell-edge user" always prefers to be co-scheduled with
cell-center users to achieve higher average SINR \cite{ZY112}.
Further consider that $\mathbf{h}_{u}$ and $\mathbf{h}_{j}$ are
Gaussian random vectors. In this special case, due to the
independence between the norm and the direction of a Gaussian vector
with \textit{i.i.d.} entries, $|\mathbf{h}_{u}|^2$ is independent
 from $\delta_{BM-1}$ because $\delta_{BM-1}$ only depends on the global
channel direction vectors of the co-scheduled users. In general case
where the users are not the ``cell-edge users'', the average SINR
can be approximated as \vspace{-0.2cm}
\begin{align} \nonumber
  \mathbf{E}\{\gamma_u^{\mathrm{C}}\} &\approx
  \frac{P\mathbf{E}\{|\mathbf{h}_{u}|^2\}\mathbf{E}\{\delta_{BM-1}\}}{M\sigma^2}
  \\ \label{E:EGC}
  &=\frac{PN_t(\sum_{i=1}^{B}\alpha_{iu}^2) \mathbf{E}\{\delta_{BM-1}\}}
  {M\sigma^2}, \vspace{-0.3cm}
\end{align}
which can serve as a lower bound of the average SINR under CoMP
transmission. The second equality in (\ref{E:EGC}) comes from the
fact that $\mathbf{E}\{|\mathbf{h}_{u}|^2\} =
\sum_{i=1}^{B}\alpha_{iu}^2\mathbf{E}\{|\mathbf{g}_{iu}|^2\}=N_t\sum_{i=1}^{B}\alpha_{iu}^2$.

\subsubsection{Average SINR Under Non-CoMP Transmission}

For a fair comparison with CoMP transmission, we consider that each
BS equally allocates its transmit power to its $M$ active users,
i.e., \vspace{-0.15cm}
\begin{equation}\label{E-PNC}
  p_{u}^{\mathrm{NC}} = \frac{P}{M\left[ (\mathbf{G}_b^H \mathbf{G}_b)^{-1}
  \right]_{1,1}}. \vspace{-0.3cm}
\end{equation}

Define
$\theta_u^{\mathrm{NC}}\triangleq\angle(\mathbf{g}_{bu},\overline{\mathbf{G}}_{b})$
as the angle between the channel of user $u$ and a subspace spanned
by the channels of its $M-1$ co-scheduled users under Non-CoMP
transmission, where
$\overline{\mathbf{G}}_b=[\mathbf{g}_{bs_1},\dots,\mathbf{g}_{bs_{M-1}}]$.
Then the received SINR of user $u$ can be obtained from
(\ref{E:GammaNC}) and (\ref{E-PNC}) as \vspace{-0.15cm}
\begin{align}  \nonumber
   \gamma_{u}^{\mathrm{NC}} &=
   \frac{P\alpha_{bu}^2} {M\left[ (\mathbf{G}_b^H \mathbf{G}_b)^{-1}
   \right]_{1,1}(P\sum_{i\neq b}^{B}\alpha_{iu}^2
  +\sigma^2)} \\ \label{E:GncEPA}
   &=  \frac{P\alpha_{bu}^2|\mathbf{g}_{bu}|^2
   \lambda_{M-1}}{M(P\sum_{i\neq b}^{B}\alpha_{iu}^2
  +\sigma^2)}, \vspace{-0.3cm}
\end{align}
where $\lambda_{M-1}\triangleq\sin^2(\theta_u^{\mathrm{NC}})$.
Similar to $\delta_{BM-1}$ defined for CoMP transmission,
$\lambda_{M-1}$ reflects the orthogonality between user $u$ and its
$M-1$ co-scheduled users.

Since under Non-CoMP the channel vector of each user is a Gaussian
random vector with \textit{i.i.d.} entries, $|\mathbf{g}_{bu}|^2$
and $\lambda_{M-1}$ are mutually independent, i.e., $
\mathbf{E}\{|\mathbf{g}_{bu}|^2\lambda_{M-1}\}=
\mathbf{E}\{|\mathbf{g}_{bu}|^2\}\mathbf{E}\{\lambda_{M-1}\} =
N_t\mathbf{E}\{\lambda_{M-1}\}$. Then the average SINR of user $u$
is \vspace{-0.2cm}
\begin{equation} \label{E:EGNC}
   \mathbf{E}\{\gamma_{u}^{\mathrm{NC}}\} =
   \frac{PN_t\alpha_{bu}^2\mathbf{E}\{\lambda_{M-1}\}}{M(P\sum_{i\neq
   b}^{B}\alpha_{iu}^2+\sigma^2)}. \vspace{-0.1cm}
\end{equation}

\subsubsection{Decision Rule of the Transmission Mode Selection}
After substituting $\mathbf{E}\{\gamma_u^{\mathrm{C}}\}$ and
$\mathbf{E}\{\gamma_u^{\mathrm{NC}}\}$ in (\ref{E:EGC}) and
(\ref{E:EGNC}), the rule in (\ref{E:LScompare}) to select
transmission mode for user $u$ turns into \vspace{-0.05cm}
\begin{equation}
  \frac{\sum_{i=1}^{B}\alpha_{iu}^2}{\alpha_{bu}^2}\cdot
  \frac{P\sum_{i\neq b}^{B}\alpha_{iu}^2+\sigma^2}{\sigma^2}>
  \frac{{\eta}^{\mathrm{NC}}}{{\eta}^{\mathrm{C}}}
  \cdot\frac{\mathbf{E}\{\lambda_{M-1}\}}{\mathbf{E}\{\delta_{BM-1}\}}, \vspace{-0.1cm}
\end{equation}
which can be rewritten as in (\ref{E:Metric}).

\begin{figure*}[!t]
\vspace{-0.2cm}
\begin{spacing}{0.9}
\begin{equation} \label{E:Metric}
  \underbrace{\left(1+\underbrace{{\frac{\sum_{i\neq b}^{B}\alpha_{iu}^2}{\alpha_{bu}^2}}}_{(a)}\right)
  \left(1+\underbrace{{\frac{P\sum_{i\neq b}^{B}\alpha_{iu}^2}{\sigma^2}}}_{(b)}\right)}_{\mathrm{Decision Variable}}
  >
  \underbrace{\frac{{\eta}^{\mathrm{NC}}}{{\eta}^{\mathrm{C}}}
  \cdot\frac{\mathbf{E}\{\lambda_{M-1}\}}{\mathbf{E}\{\delta_{BM-1}\}}}_{\mathrm{Decision
  Threshold}} \triangleq T_{\eta} \cdot T_{o} \vspace{-0.3cm}
\end{equation}
\end{spacing}
\vspace{0.3cm}\hrulefill \vspace{-0.4cm}
\end{figure*}

The value of
$T_o=\frac{\mathbf{E}\{\lambda_{M-1}\}}{\mathbf{E}\{\delta_{BM-1}\}}$
in the decision threshold depends on the scheduling method. With
random user-scheduling, it was obtained in \cite{ZY112} that
$\mathbf{E}\{\lambda_{M-1}\}=\frac{N_t-M+1}{N_t}$ and
$\mathbf{E}\{\delta_{BM-1}\}=\frac{BN_t-BM+1}{BN_t}$, then
$T_o=\frac{B(N_t-M)+B}{B(N_t-M)+1}$. When other scheduling methods
such as semi-orthogonal user scheduling (SUS) \cite{Yoo06-SUS} is
applied and the number of candidate users is large, both the values
of $\mathbf{E}\{\lambda_{M-1}\}$ and $\mathbf{E}\{\lambda_{M-1}\}$
approach 1, then $T_o \approx 1$. The value of
$T_{\eta}=\frac{{\eta}^{\mathrm{NC}}}{{\eta}^{\mathrm{C}}}$ also
relates to the scheduling. This can be seen from the expressions of
$\eta^{\mathrm{C}}$ and $\eta^{\mathrm{NC}}$ in (\ref{E:eta1}) and
the expressions of $\mathbf{E}\{\gamma_{u}^{\mathrm{C}}\}$ and
$\mathbf{E}\{\gamma_{u}^{\mathrm{NC}}\}$ in (\ref{E:EGC}) and
(\ref{E:EGNC}). When there are large number of candidate users, the
selected users are orthogonal in high probability. In this case,
$\lambda_{M-1}\approx 1$ and $\delta_{BM-1}\approx 1$, then we have \vspace{-0.1cm}
\begin{equation} \label{E:etaappro}
  {\eta}^{\mathrm{C}} \approx \frac{(\overline{\gamma}_u^{\mathrm{C}}+1)^{1-v^{\mathbf{C}}}-1}
  {\overline{\gamma}_u^{\mathrm{C}}} \ \ \mathrm{and} \ \
  {\eta}^{\mathrm{NC}} \approx \frac{(\overline{\gamma}_u^{\mathrm{NC}}+1)^{1-v^{\mathbf{NC}}}-1}
  {\overline{\gamma}_u^{\mathrm{NC}}}, \vspace{-0.05cm}
\end{equation}
where $
  \overline{\gamma}_u^{\mathrm{C}} \triangleq
  \mathbf{E}\{\gamma_{u}^{\mathrm{C}}|_{\delta_{BM-1}\approx 1}\}
  \approx\frac{PN_t(\sum_{i=1}^{B}\alpha_{iu}^2)}{M\sigma^2}$, and $
  \overline{\gamma}_u^{\mathrm{NC}} \triangleq
  \mathbf{E}\{\gamma_{u}^{\mathrm{NC}}|_{\lambda_{M-1}\approx1}\}  \approx
  \frac{PN_t\alpha_{bu}^2}{M(P\sum_{i\neq
  b}^{B}\alpha_{iu}^2+\sigma^2)}$ are respectively obtained from (\ref{E:EGC}) and (\ref{E:EGNC})
by setting $\lambda_{M-1}\approx 1$ and $\delta_{BM-1}\approx 1$.
This implies that the threshold is approximately independent from the specific scheduling method
when the number of candidate users is large. Note that the value of $T_{\eta}$ also depends on
the number of co-scheduled users. We will show by numerical results
later that the impact of the user number is minor.

The term (a) in the decision variable is in fact a ratio of the
interference to noise ratio (INR) to SNR (i.e., the reciprocal of
the signal to interference ratio) and the term (b) is the INR of
user $u$ under Non-CoMP mode. Therefore, the proposed transmission
mode selection can divide users into two groups of CoMP and Non-CoMP
users with a pre-determined decision threshold, and the result
depends on the system configuration and user location. This seems
similar to the distance threshold proposed in \cite{ZhangJun2009}
and to the soft handover region (SHO) mentioned in \cite{Tolli08}.
Nonetheless, we explicitly reveal the dependence of the threshold on
the user location and the systems parameters. Contrarily, in
\cite{ZhangJun2009} a coordinate distance is found via simulation to
maximize an effective sum rate, and in \cite{Tolli08} the metric to
determine SHO is essentially the term (a) in (\ref{E:Metric}), which
reflects the imbalance of the average channel gains of user $u$. As
will be shown later, the term (b) dominates the decision variable.
This suggests that each user can simply employ INR to determine its
transmission mode, especially for high cell-edge SNR where CoMP is
more desirable.

When the proposed decision rule is applied, each user decides its
transmission mode based on its average channel gains from multiple
BSs, and then conveys one bit to indicate its preferred mode to its
serving BS. After collecting the preference from all users, the CU
selects co-scheduled users separately from the two user groups with
a fast scheduler to achieve high throughput and ensure fairness
among the users. Such a distributed and semi-dynamic mode selection
at the user side has low signalling overhead between the users and
the BSs compared with that at the BS side. \vspace{-0.05cm}

\section{Simulation and Numerical Results}
In this section, we verify our analysis and evaluate the performance
of the proposed mode selection scheme by comparing with CoMP and
Non-CoMP systems from simulation and numerical results.

A cooperative cluster of 3 hexagonal cells is considered with a cell
radius $R=250$ m. The path loss
$\alpha_{i_u}^2=\alpha_0^2(R/d_{iu})^{\tau}$, where $\alpha_0^2$ is
the path loss at the distance $R$, $d_{iu}$ is the distance between
user $u$ and BS $i$, and $\tau=3.76$ is the path loss factor. The
cell-edge SNR is defined as the received SNR of the user located at
the distance $R$ from the BS, where the inter-cluster interference
is included and regarded as white noise. Ten users are randomly
placed in each cell. Suppose that a coherence block contains 500
channel uses ($C=500$, e.g., $T_c\approx 1.83$ ms and $W_c\approx
274$ kHz, corresponding to a 50 km/h user speed in an urban
macro-cell scenario of 3GPP \cite{3GPP-SCM}). Then the optimal
numbers of the channel uses for common and dedicated pilots can be
computed as $C_c=9.69$ and $C_d=24.2$ respectively from the results
in \cite{Caire11} and \cite{JL10}. Because there are no available
results for the overhead of CoMP in literature, we set $\beta=1$ and
$\epsilon=1$, then $v^{\mathbf{NC}}=12.59$\% and
$v^{\mathbf{C}}=28.09$\%. Unless otherwise specified, we employ this
overhead in the sequel, and we consider $M=2,N_t=4$. \vspace{-0.2cm}
\begin{figure} [!hbt]
  \centering  {
    \includegraphics[width=9.0cm]{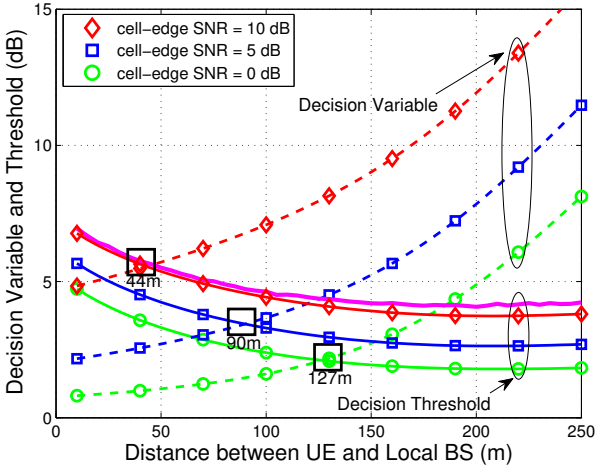}} \vspace{-0.2cm}
  \caption{Values of the decision variable and decision threshold obtained by
  numerical and simulation results,
  he solid curve without marker is the simulated decision threshold, where the SUS scheduler
  \cite{Yoo06-SUS} is used, $K$ = 10 and the cell-edge SNR = 10 dB. }
  \label{F:Decision} \vspace{-0.2cm}
\end{figure}
Figure \ref{F:Decision} shows the values of the decision variable
and the threshold provided in (\ref{E:Metric}). The border between
cell-center and cell-edge region, where the decision variable equals
the decision threshold, is also marked. We can find that the mode
selection results largely depend on the cell-edge SNR. As the SNR
grows, the CoMP region increases, i.e., CoMP transmission is more
desirable for a system with higher cell-edge SNR. To show that the
threshold is independent from the scheduling methods even with
finite number of candidate users, we also provide a simulated result
where the value of $T_o$ in the threshold is obtained by simulation
with SUS. \vspace{-0.2cm}
\begin{figure} [!hbt]
  \centering  {
    \includegraphics[width=8.0cm]{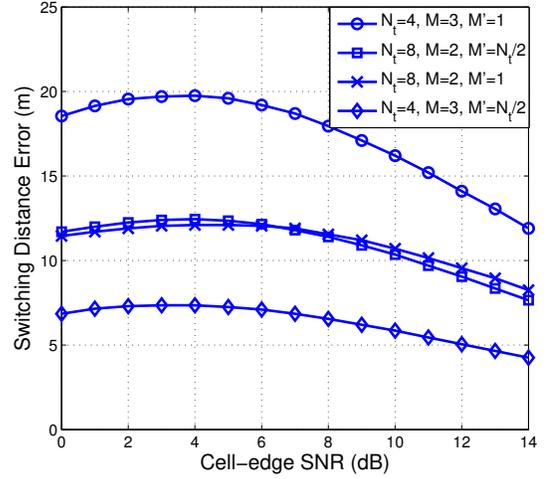}} \vspace{-0.1cm}
  \caption{Decision threshold error with estimated number of the scheduled users.}   \centering
  \label{F:Dist} \vspace{-0.2cm}
\end{figure}
\begin{figure} [!hbt]
  \centering  {
    \includegraphics[width=8.0cm]{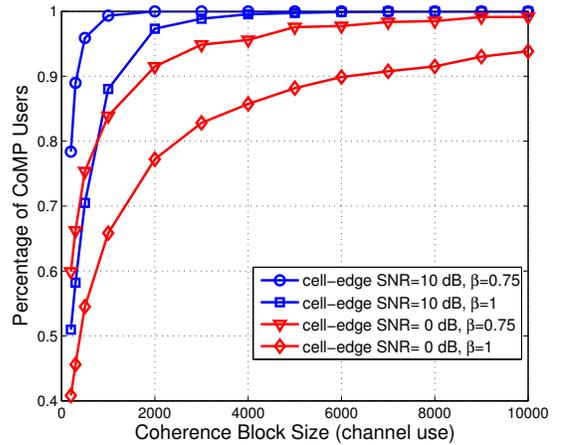}} \vspace{-0.2cm}
  \caption{Percentage of CoMP users.}
  \label{F:CoMPuser} \vspace{-0.2cm}
\end{figure}

\begin{table*}[!t]
\vspace{-0.2cm}
\begin{spacing}{0.9}
\centering \caption{Dominating factors in the decision variable
obtained from (\ref{E:Metric})}
  \begin{tabular}{c||ccccc}
  \hline
  {\small Cell-edge SNR} & {\small Switching Distance}& {\small  Percentage of CoMP-users} & {\small Term (a)}
  & {\small Term (b)} & {\small Decision Variable} \\ \hline
  10 dB & 44 m & 71\% & 0.02 dB & 5.53 dB & 5.55 dB \\
  5 dB & 90 m & 63\% & 0.13 dB& 3.31 dB&  3.44 dB \\
  0 dB & 127 m & 54\% & 0.42 dB& 1.69 dB& 2.11 dB \\
  \hline
  \end{tabular}
\end{spacing}
 \vspace{-0.2cm}
\end{table*}

In Table \Rmnum{1}, the mode switching distance under 10, 5, 0 dB
cell-edge SNRs are listed, which are computed from (\ref{E:Metric}).
We also list the corresponding term (a) and term (b) in
(\ref{E:Metric}). We can see that term (b), i.e., the INR, dominates
the decision variable, especially for high cell-edge SNR.

Remember that $T_{\eta}$ in the threshold depends on the
co-scheduled user number, $M$, but each user does not know how many
users will be scheduled. In practice, we can roughly estimate the
co-scheduled user number as $M'=N_t/2$ or even simply as $M'=1$. In
Fig. \ref{F:Dist}, we present the switching distance error caused by
the estimated $M$, which is numerically obtained from
(\ref{E:Metric}). It shows that the error is up to 20 m in various
settings. \vspace{-0.05cm}

To demonstrate the impact of the overhead on the mode switching,
Fig. \ref{F:CoMPuser} shows the percentage of CoMP users versus
coherence block size obtained by simulation. A large coherence block
size corresponds to lower speed users for a given channel delay
spread, which implies a less training overhead. The parameter $\beta
= 0.75$ reflects the fact that the CoMP system is expected to employ
sparser common pilots than Non-CoMP system. When $\beta = 1$, the
CoMP system employs common pilots with the same density as Non-CoMP
system, and the overall downlink overhead is 2.2 times over
Non-CoMP, i.e., ${v^{\mathrm{C}}}/{v^{\mathrm{NC}}}=2.2$. According
to the spatial channel model (SCM) in 3GPP \cite{3GPP-RS2}, the
coherence time $T_c$ ranges from 0.76 ms to 30.5 ms and the
coherence bandwidth $W_c$ is from 308 kHz to 1177 kHz, which
corresponds to the coherence block size ranging from 234 to 35899
channel uses. We can see that the number of CoMP users decreases
sharply with the increase of the overhead, especially when the
cell-edge SNR is high.
\begin{figure} [!hbt]
  \centering  {
    \includegraphics[width=9.0cm]{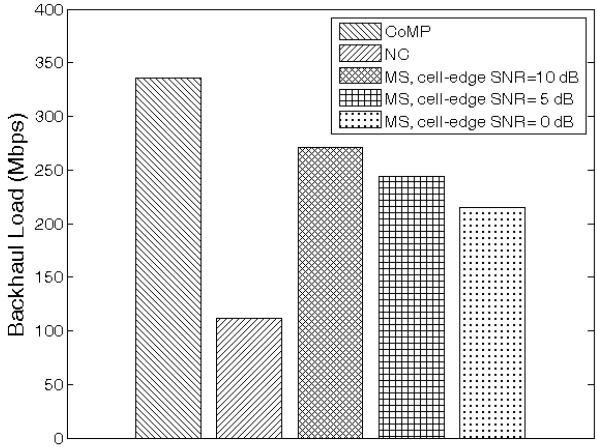}}
  \caption{Backhaul load comparison: the legend ``CoMP" denotes a system where all users are served by CoMP;
  the legend ``NC" denotes a system where all users are served by Non-CoMP,
  and ``MS" denotes the system with the proposed transmission mode selection.}
  \label{F:BHL} \vspace{-0.2cm}
\end{figure}

\begin{figure} [!hbt]
  \subfigure[Cell-edge SNR = 5 dB: the legend ``MS (Accurate $T_{\eta})$" denotes mode selection
  based on accurate $T_{\eta}$ without the approximations in (\ref{E:etaappro}), the legend ``MS (Proposed)" is the proposed mode selection method.]{
    \label{F:CDF:a} 
    \includegraphics[width=8.5cm]{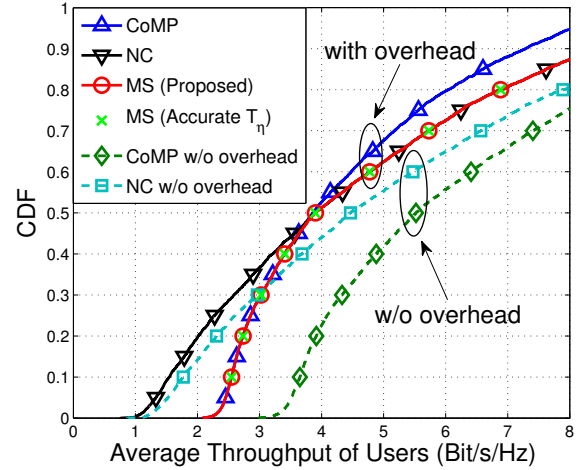}}
  \hspace{0.1cm}
  \subfigure[Cell-edge SNR = 0 dB: the legend ``MS (Simulated)" denotes mode selection based on the simulated average throughputs rather than the throughput
  upper bounds in (\ref{E:rateAppro}) and
  (\ref{E:rateAppro1}).]{
    \label{F:CDF:b} 
    \includegraphics[width=8.5cm]{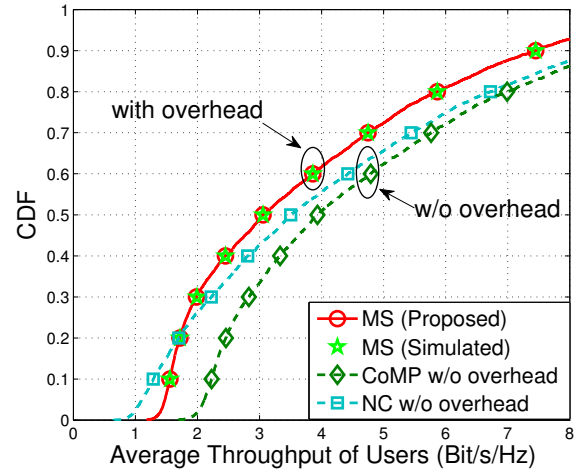}}
  \caption{Average throughput of users: ``CoMP w/o overhead" and ``NC w/o overhead respectively denote the
  throughputs of the systems
  where all users are served by CoMP and by Non-CoMP, and the overhead is not considered.}   \centering
  \label{F:CDF} \vspace{-0.2cm}
\end{figure}

In addition to mitigating the adverse impact of the overhead, the
mode switching also leads to a backhaul load reduction by reducing
the number of CoMP users. In Fig. \ref{F:BHL}, we compare the
backhaul loads of three systems, which are the average data rate
from the CU to each BS, where the load for sharing CSI is ignored
\cite{Huang2009}. We consider an orthogonal frequency division
multiplexing (OFDM) system with $W = 20$ MHz bandwidth that is
divided into 1024 subcarriers, and the symbol duration is 71 us. 16
QAM modulation is employed, thereby the data rate of each user is
$C_s = 4\ \mathrm{bit}/71\ \mathrm{us}/20\ \mathrm{KHz} = 2.8$
bps/Hz. Under Non-CoMP each BS only needs the data for the $M$ local
users, hence the backhaul load is $MC_sW$ bps. Under CoMP the data
for all the $BM$ scheduled users should be available at each BS,
thus the load is $BMC_sW$ bps. By selecting transmission mode
between the two modes, the backhaul load is $(p_cB+1-p_c)MC_sW$ bps,
where $p_c$ denotes the percentage of CoMP users. We can see from
the simulation results that the backhaul load after mode selection
is about half of that under CoMP system in 0 dB cell-edge SNR.
\vspace{-0.05cm}

In Fig. \ref{F:CDF}, we simulate the average throughput of the
proposed mode selection method and compare it with those of pure
CoMP and Non-CoMP transmission. The results are obtained through
1000 random trials, where in each trial the throughput of the
selected users are averaged over 100 \textit{i.i.d.} Rayleigh flat
fading channels. SUS algorithm is applied for scheduling. As
expected, CoMP outperforms Non-CoMP when the overhead is not
considered. However, after taking into account the overhead, only
cell-edge users have performance gain under CoMP transmission while
cell-center users suffer from performance loss compared with
Non-CoMP transmission. The proposed transmission mode selection can
adapt to the user location and cell-edge SNR, thus achieves good
performance for both cell-edge and cell-center users. We can see
from Fig. \ref{F:CDF:a} that the approximation on
$\eta^{\mathrm{NC}}$ and $\eta^{\mathrm{C}}$ in (\ref{E:etaappro})
almost has no impact on the performance. In Fig. \ref{F:CDF:b}, we
provide the performance of the mode selection schemes under a
different cell-edge SNR when the mode is selected based on the
simulated average throughput and on the throughput upper bounds in
(\ref{E:rateAppro}) and (\ref{E:rateAppro1}). The results of these
two schemes are overlapped, which indicates that the upper bounds
are very tight. To make the figure more clear, we do not show the
performance for CoMP and Non-CoMP, which is similar to those in Fig.
\ref{F:CDF:a}. \vspace{-0.1cm}

\section{Conclusions}
We proposed a semi-dynamic transmission mode selection method
between CoMP-JP and Non- CoMP, aiming at maximizing the downlink
throughput after accounting for the training overhead. The decision
rule is in closed-form, which has an explicit relationship with the
average channel gains of each user and various system parameters.
Simulation results showed that each user should employ INR rather
than simply use channel gain imbalance to decide its transmission
mode especially at high cell-edge SNR, where most of the users
moving with moderate speed prefer CoMP transmission. \vspace{-0.2cm}

\section*{Acknowledgement}
The authors would like to thank Prof. Andreas F. Molisch for the helpful discussions.

\bibliography{IEEEabrv,Zhangbib}

\newpage

\end{document}